\documentclass{article}
\usepackage{float}
\usepackage[hidelinks]{hyperref} 
\usepackage{fancyhdr}

 \usepackage{graphicx} 

\usepackage[nonatbib,main,final]{neurips_2025}

\usepackage[backend=biber,style=ieee,maxnames=99]{biblatex}
\usepackage[utf8]{inputenc} 
\usepackage[T1]{fontenc}    
\usepackage{hyperref}       
\usepackage{url}            
\usepackage{booktabs}       
\usepackage{amsfonts}       
\usepackage{nicefrac}       
\usepackage{microtype}      
\usepackage{xcolor}         
\usepackage{booktabs}      
\usepackage{multirow}      
\usepackage{siunitx}       
\usepackage{makecell}      

\title{SoulX-Singer: Towards High-Quality Zero-Shot Singing Voice Synthesis}

\author{%
\textbf{Jiale Qian}\textsuperscript{1}\thanks{Equal contribution.} \and
\textbf{Hao Meng}\textsuperscript{1,3}\footnotemark[1] \and
\textbf{Tian Zheng}\textsuperscript{1} \and
\textbf{Pengcheng Zhu}\textsuperscript{2} \and
\textbf{Haopeng Lin}\textsuperscript{1} \and
\textbf{Yuhang Dai}\textsuperscript{1,4} \and
\textbf{Hanke Xie}\textsuperscript{1,4} \and
\textbf{Wenxiao Cao}\textsuperscript{1} \and
\textbf{Ruixuan Shang}\textsuperscript{1} \and
\textbf{Jun Wu}\textsuperscript{1} \and
\textbf{Hongmei Liu}\textsuperscript{1} \and
\textbf{Hanlin Wen}\textsuperscript{1} \and
\textbf{Jian Zhao}\textsuperscript{2} \and
\textbf{Zhonglin Jiang}\textsuperscript{2} \and
\textbf{Yong Chen}\textsuperscript{2} \and
\textbf{Shunshun Yin}\textsuperscript{1} \and
\textbf{Ming Tao}\textsuperscript{1} \and
\textbf{Jianguo Wei}\textsuperscript{3} \and
\textbf{Lei Xie}\textsuperscript{4} \and
\textbf{Xinsheng Wang}\textsuperscript{1}\thanks{Corresponding author. \texttt{qianjiale@soulapp.cn}, \texttt{menghao@soulapp.cn}, \texttt{wangxinsheng@soulapp.cn}} \and \\
\textsuperscript{1}Soul AI Lab, China \\
\textsuperscript{2}AI Center, Geely Automobile Research Institute (Ningbo) Co., Ltd., Ningbo, China \\
\textsuperscript{3}Audio-Visual Cognitive Computing Team, Tianjin University, Tianjin, China \\
\textsuperscript{4}Audio, Speech and Language Processing Group (ASLP@NPU),\\
Northwestern Polytechnical University, Xi’an, China \\
}

\begin{document}

\maketitle
\thispagestyle{fancy}

\vspace{-2.5em}

\begin{abstract}

\vspace{-0.6em}


While recent years have witnessed rapid progress in speech synthesis, open-source singing voice synthesis (SVS) systems still face significant barriers to industrial deployment, particularly in terms of robustness and zero-shot generalization. In this report, we introduce SoulX-Singer, a high-quality open-source SVS system designed with practical deployment considerations in mind. SoulX-Singer supports controllable singing generation conditioned on either symbolic musical scores (MIDI) or melodic representations, enabling flexible and expressive control in real-world production workflows. Trained on more than 42,000 hours of vocal data, the system supports Mandarin Chinese, English, and Cantonese and consistently achieves state-of-the-art synthesis quality across languages under diverse musical conditions. Furthermore, to enable reliable evaluation of zero-shot SVS performance in practical scenarios, we construct SoulX-Singer-Eval, a dedicated benchmark with strict training-test disentanglement, facilitating systematic assessment in zero-shot settings.

\end{abstract}
\vspace{-0.8em}

\hspace*{0.09\textwidth}
\begin{minipage}{0.9\textwidth}
\small
\textbf{Demo page:} \href{https://soul-ailab.github.io/soulx-singer/}{\textcolor{blue}{https://soul-ailab.github.io/soulx-singer}} \\
\textbf{Source code:} \href{https://github.com/Soul-AILab/SoulX-Singer}{\textcolor{blue}{https://github.com/Soul-AILab/SoulX-Singer}}
\end{minipage}

\begin{figure}[hb]
    \centering
    \includegraphics[width=.92\linewidth]{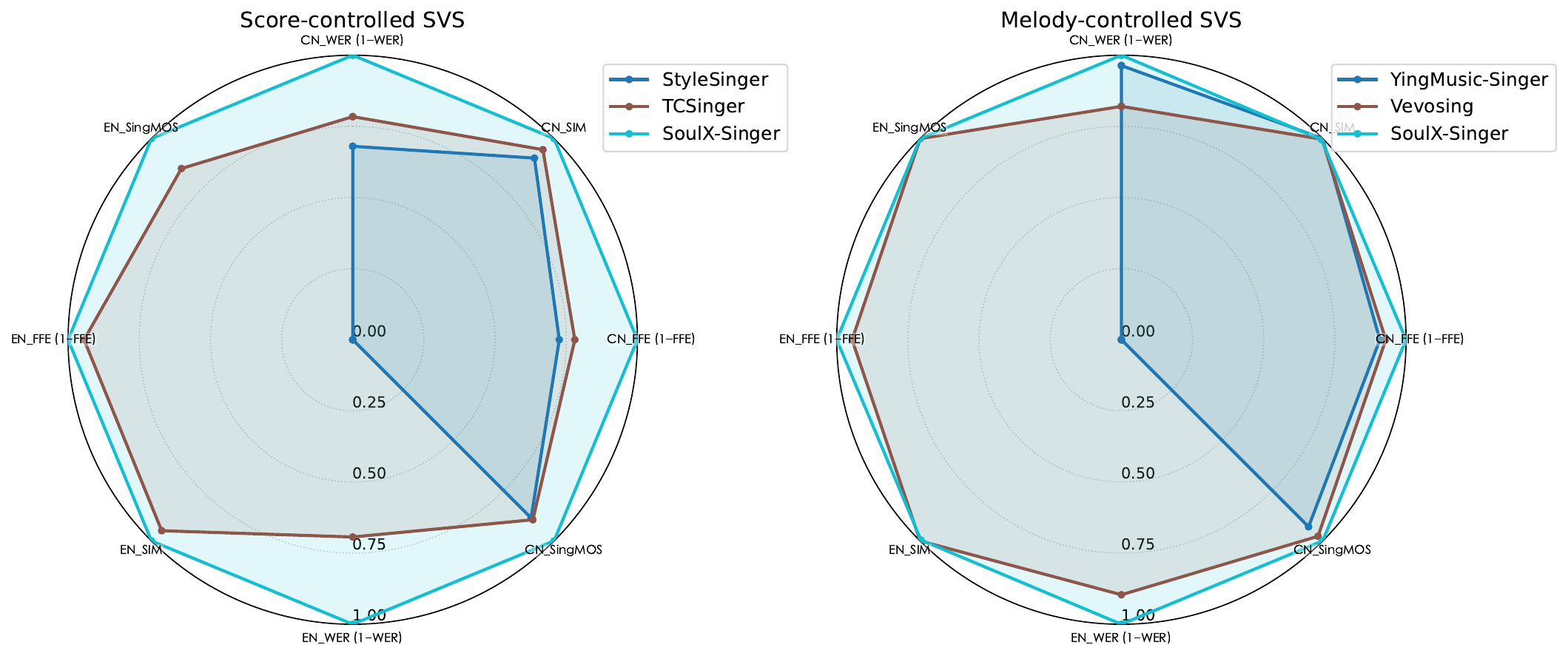}
    \caption{Performance of SoulX-Singer.}
    \label{fig:soulxpodcast}
\end{figure}

\section{Introduction}

Singing Voice Synthesis (SVS) aims to generate expressive human vocals conditioned on lyrics and musical scores. Although significant progress has recently been made in both speech synthesis~\cite{chen2025f5tts, wang2025spark, zhou2025indextts2, du2025cosyvoice, xie2025fireredtts, zhou2025voxcpm, hu2026qwen3} and music generation~\cite{jiang2025diffrhythm, yuan2025yue, yang2025songbloom}, high-quality singing voice generation with flexible control in a zero-shot setting remains largely unavailable. To address this gap, we introduce SoulX-Singer, a singing voice synthesis system capable of high-quality zero-shot generation across multiple languages.

Early work on SVS primarily focused on synthesizing singing voices from speakers or singers observed during training~\cite{liu2022diffsinger, xue2022learn2sing, lei2023unisyn}. Representative systems such as DiffSinger~\cite{liu2022diffsinger} were trained on relatively small-scale datasets with carefully curated annotations, and therefore lacked the ability to generalize to unseen singers. This limitation significantly restricts the applicability of such models in real-world scenarios. Subsequent studies, including StyleSinger~\cite{zhang2024stylesinger} and the TCSinger series~\cite{zhang2024tcsinger, zhang2025tcsinger2}, began to explore zero-shot SVS. However, these methods were still trained on datasets consisting of only a few hundred hours of singing data from a limited number of singers, making it difficult to achieve robust zero-shot generalization in practice.

More recently, Vevo2~\cite{zhang2025vevo2} and YingMusic-Singer~\cite{zheng2025yingmusic} leveraged the scaling capabilities of the Transformer~\cite{vaswani2017attention} and Diffusion Transformer (DiT)~\cite{dhariwal2021diffusion} architecture, respectively, expanding singing datasets to the scale of several thousand hours. Despite these advances, both systems adopt a melody-driven synthesis paradigm and do not provide note-level duration control, which leads to two key limitations. First, melody extraction from existing songs is required, preventing song generation purely from musical scores and lyrics. Second, the absence of explicit note duration modeling makes syllable-level timing uncontrollable, resulting in temporal misalignment between the synthesized vocals and the original accompaniment. This severely limits practical usage in music production workflows, particularly for mixing and arrangement.

To address these challenges, we propose \textbf{SoulX-Singer}. In contrast to prior work relying on at most several thousand hours of training data, we construct a large-scale vocal dataset comprising over \textbf{42,000 hours} of singing audio, substantially enhancing the model’s zero-shot generalization capability. From an algorithmic perspective, SoulX-Singer is designed to support both \textbf{score-based} controllable singing generation and \textbf{melody-conditioned} synthesis, enabling flexible control under diverse real-world scenarios. This dual-control design allows SoulX-Singer to better accommodate music creation workflows based on symbolic musical scores as well as scenarios requiring melody-guided generation from existing songs.

The main contributions of this work can be summarized as follows:

\begin{itemize}
    \item \textbf{A High-Quality Zero-shot SVS Model}. We propose \textbf{SoulX-Singer}, a high-quality singing voice synthesis model designed for zero-shot scenarios. SoulX-Singer supports both \textbf{music score-based inputs} and \textbf{melody-based inputs} within a unified framework, enabling high-fidelity timbre cloning, reference-based singing style transfer, and flexible editing of both musical scores and lyrics. This design makes the system suitable for a wide range of real-world music production workflows.

    \item \textbf{A Large-Scale Singing Voice Data Processing Pipeline}. We introduce a large-scale data processing pipeline that takes songs with background music as input and automatically produces clean vocal recordings paired with aligned lyrics and musical scores. Using this pipeline, we construct a multilingual singing dataset comprising over 42,000 hours of vocal recordings in \textbf{Mandarin Chinese}, \textbf{English}, and \textbf{Cantonese}, representing an order-of-magnitude increase compared to the data scale used in existing SVS studies.

    \item \textbf{A Dedicated Evaluation Benchmark for Zero-shot SVS}. To facilitate systematic and reproducible evaluation of zero-shot SVS performance, we introduce a dedicated benchmark dataset featuring \textbf{50 unseen singers} in Mandarin Chinese and English, accompanied by \textbf{fine-grained, note-level score annotations}. This benchmark provides a standardized evaluation protocol for assessing synthesis quality, controllability, and generalization, and serves as a reliable testbed for future SVS research.
    
\end{itemize}

In the following sections, we introduce the data construction and model architecture of SoulX-Singer, followed by a comprehensive evaluation of its performance on singing voice synthesis tasks.

\section{Method}

Clean vocal recordings annotated with aligned MIDI scores and lyrics are essential for training SVS models with fine-grained, score-based controllability. In this section, we first describe the data processing pipeline adopted in this work, including vocal extraction from mixed songs and the annotation of MIDI information and lyric transcriptions. We then present the overall architecture and core algorithms of SoulX-Singer.

\subsection{Data Processing}

\begin{figure}[h]
    \centering
    \includegraphics[width=0.9\linewidth]{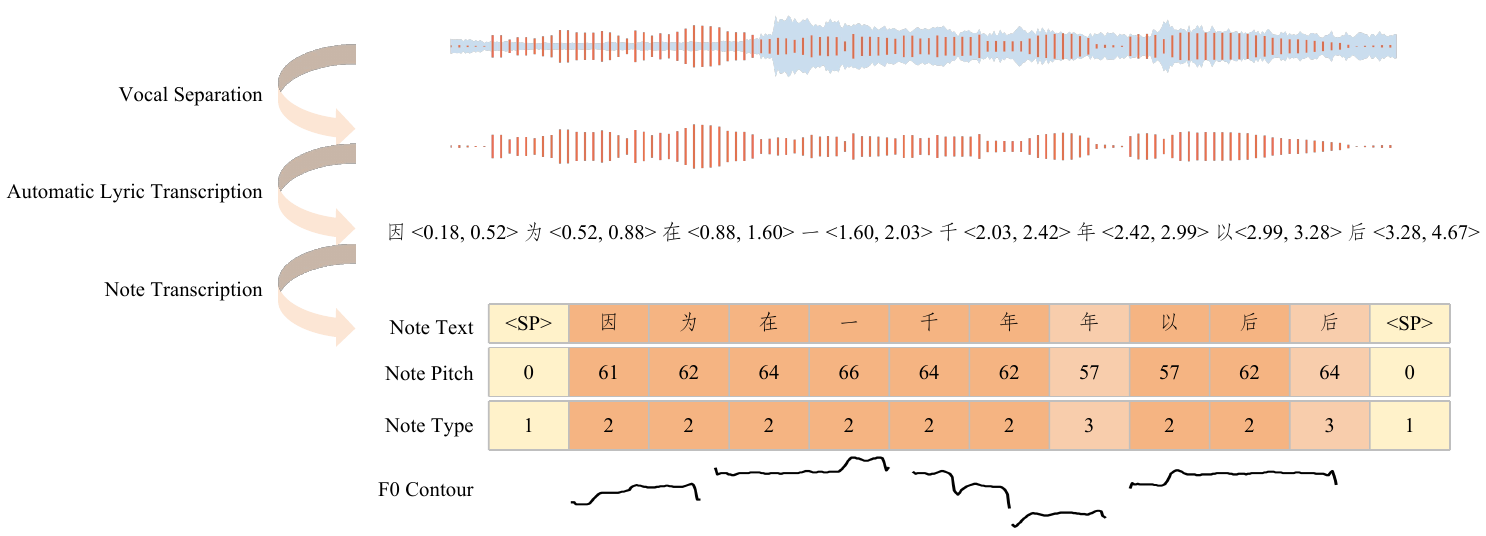}
    \caption{Pipeline for large-scale singing data curation: from raw audio extraction to time-aligned MIDI and text formulation.}
    \label{fig:data_pipeline}
\end{figure}

To obtain aligned vocal audio, MIDI, and lyric annotations, as illustrated in Figure~\ref{fig:data_pipeline}, the overall data processing workflow consists of vocal separation, lyric transcription, and note transcription. In addition, to enable melody-controlled SVS, the fundamental frequency (F0) is extracted from the vocal recordings.



\subsubsection{Processing Workflow}

\textbf{Vocal Separation}. 
To obtain high-purity dry vocals devoid of backing harmonies, we employ a two-stage vocal extraction process. Specifically, a pretrained lead vocal separation model\footnote{\url{https://huggingface.co/becruily/mel-band-roformer-karaoke}}
 is adopted to isolate the lead vocal and suppress accompanying harmonies. Since commercial recordings often contain reverberation introduced during mixing, a pretrained vocal de-reverberation model\footnote{\url{https://huggingface.co/anvuew/dereverb_mel_band_roformer}}
 is subsequently applied. Both models are based on Mel-Band Roformer \cite{wang2023mel}, providing a clean acoustic foundation that supports high-fidelity synthesis and large-scale training.




\textbf{Automatic Lyric Transcription}. 
To ensure accurate alignment of lyrics with vocals, we first perform robust language identification. Specifically, SenseVoiceSmall\footnote{\url{https://www.modelscope.cn/models/iic/SenseVoiceSmall}}~\cite{an2024funaudiollm} is adopted as a pretrained backbone and fine-tuned on a singing dataset labeled with Mandarin, Cantonese, and English. Once the language is determined, we extract lyrics and word-level timestamps using language-specific ASR models: Paraformer\footnote{\url{https://modelscope.cn/models/iic/speech_seaco_paraformer_large_asr_nat-zh-cn-16k-common-vocab8404-pytorch}}~\cite{gao22paraformer} for Mandarin and Cantonese, and Parakeet-TDT-0.6B-V2\footnote{\url{https://huggingface.co/nvidia/parakeet-tdt-0.6b-v2}} for English. To guarantee data quality, extracted lyrics are compared with reference lyrics annotated at the sentence level. Samples containing insertion or deletion errors are discarded, and substitution errors are corrected using the reference words. This procedure ensures the linguistic accuracy required for robust large-scale training.

\textbf{Note Transcription}. 
To generate note-level representations aligned with the lyrics, we employ the ROSVOT model~\cite{li2024rosvot} for pitch-class estimation and note boundary detection. Specifically, timestamped lyrics from the previous stage are used as input, producing a sequence of note-level tokens—including text, pitch, note type, and duration—precisely aligned along the temporal axis. These aligned tokens provide the necessary foundation for controllable and expressive singing voice synthesis.

\begin{figure}[t]
    \centering
    \includegraphics[width=0.9\linewidth]{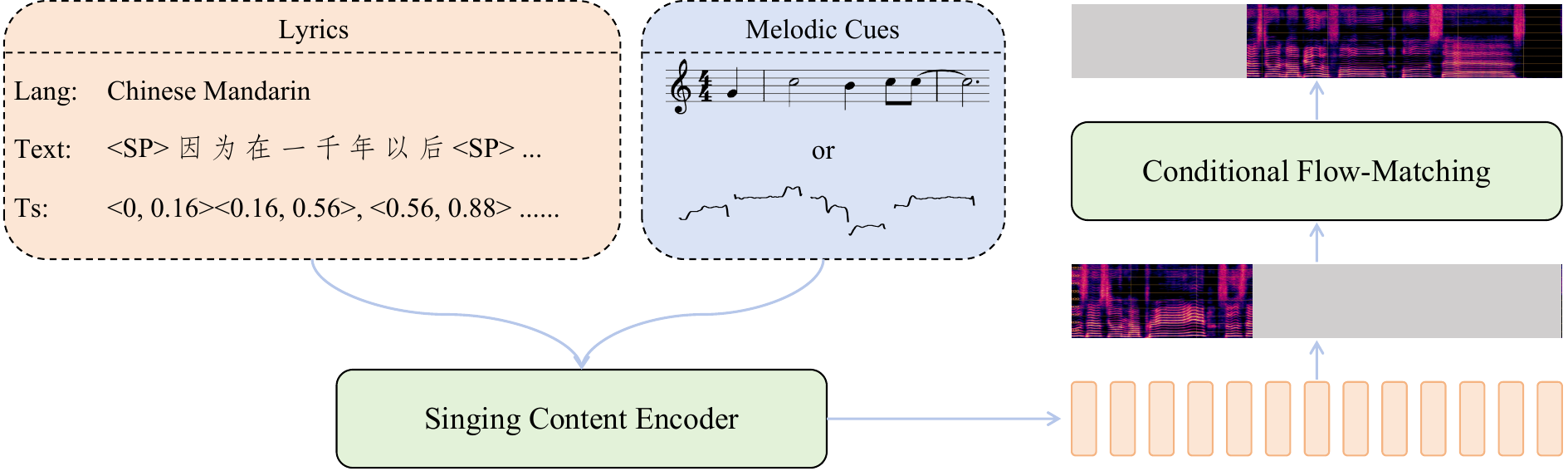}
    \caption{Overview of SoulX-Singer.}
    \label{fig:architecture}
\end{figure}

\subsubsection{Corpus Overview}

Using the aforementioned processing methods, we obtained approximately 42,000 hours of high-quality vocal recordings, with roughly 20k hours each in Mandarin and English, and about 2k hours in Cantonese. To enable precise modeling of singing voice, the dataset is organized at the musical note level, which serves as the fundamental unit for synthesis. Specifically, each note is represented as a tuple consisting of its corresponding textual token, pitch class, and note type. The note type is defined as a categorical attribute, with values 1, 2, and 3 corresponding to rest, lyric, and slur notes, respectively. This note-level representation ensures high-precision alignment between linguistic and melodic structures, providing a robust foundation for controllable and expressive singing voice synthesis.

\subsection{SoulX-Singer}

SoulX-Singer is a non-autoregressive (NAR) singing voice synthesis model based on flow matching, designed for high-quality and controllable singing voice generation. The core backbone of the model is a flow matching decoder built with Diffusion Transformer (DiT)~\cite{peebles2023scalable}, which takes lyrics and melody cues as input and predicts mel-spectrograms. The predicted mel-spectrograms are subsequently converted into waveforms using a neural vocoder.

To effectively encode the diverse and multimodal information required for SVS, i.e, lyrics, musical scores, note types, and F0, SoulX-Singer incorporates a Singing Content Encode. This encoder transforms the input features into rich, temporally aligned representations, providing the decoder with a structured and informative latent space for accurate mel-spectrogram generation. By combining flow-based NAR modeling with specialized content encoding, SoulX-Singer achieves both expressive control over vocal performance and efficient, high-fidelity synthesis.






\subsubsection{Feature Representation}

\textbf{Textual Representation}. For Mandarin and Cantonese, the modeling units are defined as character-level pinyin, whereas English is represented using phonemes. To explicitly distinguish phoneme boundaries across different words, the phoneme sequence corresponding to each English word is wrapped with special boundary tokens, starting with <BOW> and ending with <EOW>. To further disambiguate pinyin representations between Mandarin and Cantonese, language-specific tags are appended to the pinyin tokens, enabling the model to distinguish pronunciation patterns across languages. The text embeddings are obtained via a linear embedding layer.

\textbf{Melodic Representation}. The melodic input consists of discrete note pitch sequences and continuous F0 sequences. Both sequences are first processed through binary gating layers, followed by a linear projection to produce note pitch embeddings and F0 embeddings, respectively. The gating mechanism regulates the contribution of each prosodic feature: during training, either the note pitch or F0 input is randomly dropped to encourage robust feature extraction from a single modality; during inference, the model can flexibly enable the corresponding gate to perform melody-based or score-based generation.

\textbf{Length Regulation and Feature Integration}. To unify heterogeneous representations into a shared temporal resolution before decoding, we adopt a length regulator as the core feature integration mechanism. Specifically, the embeddings corresponding to note type, note pitch, and text tokens are expanded according to the duration of each musical note, aligning all note-level representations to the mel-spectrogram time scale. After length expansion, these embeddings share identical temporal and feature dimensions and are combined through element-wise addition, producing a unified conditioning sequence that is subsequently fed into the decoder. By explicitly enforcing note-to-mel alignment, the length regulator enables precise synchronization between linguistic content and melodic structure, which is essential for controllable and expressive singing voice synthesis.

\subsubsection{Training}

The training of SoulX-Singer is conducted in two stages to progressively enhance model robustness and long-form generation capability. In the first stage, the model is trained on relatively short audio segments, with durations ranging from 2 s to 16 s. During this stage, the prompt mel-spectrogram is deliberately sampled from a non-adjacent segment of the target audio. This design encourages the model to rely less on local acoustic continuity and more on the provided linguistic and musical conditions, thereby improving robustness and generalization under diverse prompt conditions.

In the second stage, the training strategy is shifted toward long-form singing voice modeling. Adjacent audio segments are concatenated to construct longer training samples with durations between 30 s and 90 s, enabling the model to capture long-range temporal dependencies in singing performances. To further strengthen the model’s prompt-following capability, the prompt audio in this stage is sampled from the immediately preceding adjacent segment of the target audio. 

By adopting this two-stage training strategy—progressing from short, non-adjacent prompting to long, contextually adjacent prompting—SoulX-Singer achieves both robust conditional generation and effective modeling of long-duration singing audio.

\subsubsection{Inference}




During inference, SoulX-Singer offers high flexibility by supporting two complementary generation modes, which can be seamlessly switched according to the available prosodic control signals.

\textbf{Melody-control Mode}. This mode is intended for scenarios where a target melody is available. In this configuration the model takes the target lyrics together with a continuous F0 contour extracted from a reference audio as primary inputs. This design enables faithful preservation of fine-grained melodic details and expressive vocal techniques, while still allowing flexible modification of lyrics or transformation of timbre.

\textbf{Score-control Mode}. This mode is designed for creative synthesis scenarios. In this mode, the inputs consist solely of MIDI score information and the target lyrics. The model autonomously predicts naturalistic acoustic features constrained by the provided score. By eliminating the dependency on pre-recorded melody lines, this mode provides creators with greater artistic freedom for high-fidelity zero-shot SVS.

\section{Performance of SoulX-Singer}

\subsection{Evaluation Dataset}

Due to the absence of a widely adopted and standardized benchmark for singing voice synthesis (SVS), we construct two complementary evaluation datasets to comprehensively assess the performance of SoulX-Singer under both open-source and zero-shot conditions: GMO-SVS and SoulX-Singer-Eval.

\textbf{GMO-SVS}. The GMO-SVS dataset is built upon multiple publicly available SVS corpora, including \textbf{G}TSinger \cite{zhang2024gtsinger}, \textbf{M}4Singer \cite{zhang2022m4singer}, and \textbf{O}pencpop \cite{wang22opencpop}. For M4Singer and Opencpop, we directly adopt their official test splits. Specifically, M4Singer contains 8 Mandarin songs performed by 4 singers spanning four vocal ranges, while Opencpop provides 5 Mandarin songs sung by a single professional female singer. GTSinger contributes 25 songs in both English and Mandarin from 5 singers, covering a diverse set of vocal techniques and expressive styles.

In total, GMO-SVS consists of 802 samples. For each song, the first sentence is used as the acoustic prompt, while the remaining content is synthesized by the evaluated models. Importantly, GMO-SVS preserves the ground-truth recordings of the prompt singers, enabling a comprehensive evaluation of pronunciation accuracy, prosodic consistency, and overall synthesis quality. It is worth emphasizing that none of the above open-source datasets are used during the training of SoulX-Singer, ensuring a fair and unbiased evaluation.

To further evaluate the model’s performance in singing voice editing scenarios, we create modified versions of the target lyrics. Specifically, the original lyrics of each song are rewritten using the DeepSeek-V large language model, while strictly maintaining the same number of words as the original. This setup allows us to assess how well the model can adapt to lyric modifications while preserving melodic and expressive characteristics.

\textbf{SoulX-Singer-Eval}. This dataset is a newly collected dataset designed to evaluate zero-shot generalization on unseen speakers. It contains 100 singing segments from 50 distinct individuals (25 Mandarin and 25 English speakers), with 2 segments per speaker. The Mandarin data were collected from recruited professional and amateur singers who consented to open-source their voice data for academic purposes. The English segments were sliced and filtered from the multitrack \textit{Mixing Secrets} dataset \cite{gururani2017mixing}. Crucially, all segments in SoulX-Singer-Eval underwent precise manual melody annotation to meet the prompt input requirements of various zero-shot SVS models. The target lyrics and melody for synthesis are randomly selected from 15 Mandarin and 15 English tracks in GMO-SVS. This set introduces speakers unseen by any baseline models, providing a rigorous benchmark for timbre cloning and style transfer capabilities.

\subsection{Evaluation Metrics}

To comprehensively assess the performance of singing voice synthesis, we evaluate SoulX-Singer across multiple key dimensions, including melodic accuracy, timbre similarity, intelligibility, and overall singing quality. Each metric is carefully selected to quantify different aspects of synthesized vocal fidelity and expressiveness.


\textbf{Melodic Accuracy}. We measure pitch precision using F0 Frame Error (FFE), defined as the proportion of frames in which the predicted pitch deviates by more than 20\% from the ground-truth F0. 


\textbf{Timbre Similarity}. To evaluate the model’s ability to reproduce the timbre of the acoustic prompt in zero-shot scenarios, we compute cosine similarity (SIM) between speaker embeddings of the synthesized audio and the corresponding prompt. The embeddings are extracted using a WavLM-based speaker verification model \cite{chen2022wavlm}.


\textbf{Intelligibility}. We assess pronunciation accuracy via Word Error Rate (WER), calculated by comparing the target lyrics with automatic speech recognition (ASR) transcriptions of the synthesized audio. For Mandarin samples, character-level error rate (CER) is the standard metric; however, for simplicity and consistency across languages, we report all results using WER. Paraformer \cite{gao22paraformer} is used for Mandarin, and Whisper-large-v3 \cite{radford2023whisper} is used for English samples.


\textbf{Singing Quality}. Overall perceptual quality is evaluated using two learned objective metrics. SingMOS \cite{tang2025singmos} is a singing-specific quality metric trained to align closely with human perception. Sheet-SSQA (Sheet) \cite{huang2024mos}, derived from the MOS-Bench framework, is used to assess perceptual quality in zero-shot scenarios and demonstrates strong generalization across unseen speakers.


\subsection{Results}


\subsubsection{Performance on the GMO-SVS}
Comparisons of SoulX-Singer and other baseline methods on GMO-SVS are presented in Table~\ref{tab:synthesis_editing}, where Singing Voice Editing indicates that the lyrics are rewritten. For the SVS task, SoulX-Singer outperforms all baseline models in both Mandarin and English. When controlled with continuous F0 contours (melody-control mode), the model achieves the lowest FFE, significantly surpassing the best baseline, YingMusic-Singer. This demonstrates that explicit acoustic features guide the flow-matching decoder to generate accurate pitch trajectories. In contrast, when driven by discrete MIDI notes (score-control mode), SoulX-Singer attains the lowest WER, outperforming Vevosing and TCSinger. This indicates that MIDI-based timing constraints help stabilize pronunciation and rhythm, particularly for complex phonemes. Across both modes, SoulX-Singer also achieves state-of-the-art performance on SingMOS and SIM, confirming the robustness and generalization capability of the architecture under both pitch- and score-conditioned synthesis scenarios.

The results under Singing Voice Editing reflect the performance of different models when the lyrics are modified. As shown in the table, melody-based methods exhibit a noticeable drop in intelligibility compared to the original lyrics. This decline arises because the original melody and lyrics are inherently correlated—for example, the pitch contours are aligned with the original words—so modifying the lyrics introduces a certain degree of mismatch. In contrast, SoulX-Singer, which supports MIDI score-based control, maintains better performance under lyric modification. By relying on the explicit score rather than the original acoustic melody, the model can effectively accommodate rewritten lyrics without sacrificing pronunciation accuracy or rhythmic consistency.

\begin{table}[htbp]
\centering
\small
\caption{Performance of different models on GMO-SVS ($\uparrow$ larger is better, $\downarrow$ smaller is better; bold indicates the best result).}
\label{tab:synthesis_editing}
\vspace{2mm}
\makebox[\textwidth][c]{
\resizebox{1.0\textwidth}{!}{
\begin{tabular}{llcccccccccc}
\toprule
\multicolumn{12}{c}{\textbf{Singing Voice Synthesis}} \\
\midrule
\multirow{2}{*}{Model} & \multirow{2}{*}{Control} & \multicolumn{5}{c}{Chinese} & \multicolumn{5}{c}{English} \\
\cmidrule(lr){3-7} \cmidrule(lr){8-12}
& & WER $\downarrow$ & SIM $\uparrow$ & FFE $\downarrow$ & SingMOS $\uparrow$ & Sheet $\uparrow$ & WER $\downarrow$ & SIM $\uparrow$ & FFE $\downarrow$ & SingMOS $\uparrow$ & Sheet $\uparrow$ \\
\midrule
GroundTruth & - & 0.074 & - & - & 4.624 & 4.334 & 0.197 & - & - & 4.441 & 3.825 \\
\midrule
StyleSinger & Score & 0.367 & 0.817 & 0.363 & 3.938 & 3.819 & - & - & - & - & - \\
TCSinger & Score & 0.270 & 0.855 & 0.315 & 3.977 & 3.843 & 0.410 & 0.879 & 0.211 & 3.662 & 3.482 \\
YingMusic-Singer & Melody & 0.099 & 0.902 & 0.132 & 4.145 & 3.620 & - & - & - & - & - \\
Vevosing & Melody & 0.233 & 0.899 & 0.112 & 4.355 & 4.005 & 0.239 & 0.922 & 0.088 & 4.321 & 3.699 \\
\midrule
SoulX-Singer & Melody & \textbf{0.065} & 0.897 & \textbf{0.044} & \textbf{4.458} & \textbf{4.110} & 0.151 & 0.918 & \textbf{0.036} & \textbf{4.323} & \textbf{3.751} \\
SoulX-Singer & Score & 0.069 & \textbf{0.905} & 0.122 & 4.445 & 4.107 & \textbf{0.149} & \textbf{0.926} & 0.164 & 4.303 & 3.705 \\
\midrule
\multicolumn{12}{c}{\textbf{Singing Voice Editing}} \\
\midrule
YingMusic-Singer & Melody & 0.146 & \textbf{0.904} & 0.180 & 4.107 & 3.612 & - & - & - & - & - \\
Vevosing & Melody & 0.308 & 0.897 & 0.148 & 4.328 & 4.011 & 0.484 & \textbf{0.924} & 0.108 & 4.267 & \textbf{3.691} \\
\midrule
SoulX-Singer & Melody & 0.212 & 0.895 & \textbf{0.057} & 4.325 & 4.024 & 0.444 & 0.911 & \textbf{0.047} & 4.141 & 3.615 \\
SoulX-Singer & Score & \textbf{0.089} & \textbf{0.904} & 0.181 & \textbf{4.447} & \textbf{4.125} & \textbf{0.213} & 0.920 & 0.200 & \textbf{4.269} & 3.678 \\
\bottomrule
\end{tabular}
}
}
\end{table}

\subsubsection{Performance on the SoulX-Singer-Eval} 

To evaluate the model’s capability in timbre cloning and style transfer, we conduct experiments on SoulX-Singer-Eval, which contains speakers that are completely unseen during training for all compared methods. As shown in Table~\ref{tab:unseen_parallel}, SoulX-Singer consistently outperforms baseline models even with unseen target speakers. The SoulX-Singer controlled with music scores achieves the highest SIM scores (0.922 for Mandarin and 0.914 for English), demonstrating robust zero-shot voice cloning. 




\begin{table}[htp]
\centering
\small
\caption{Performance of different models on SoulX-Singer-Eval ($\uparrow$ larger is better, $\downarrow$ smaller is better; bold indicates the best result).}
\label{tab:unseen_parallel}
\vspace{2mm}
\resizebox{\linewidth}{!}{
\begin{tabular}{llcccccccc}
\toprule
\multirow{2}{*}{Model} & \multirow{2}{*}{Control} & \multicolumn{4}{c}{\textbf{Chinese}} & \multicolumn{4}{c}{\textbf{English}} \\
\cmidrule(lr){3-6} \cmidrule(lr){7-10}
 & & WER $\downarrow$ & SIM $\uparrow$ & SingMOS $\uparrow$ & Sheet $\uparrow$ & WER $\downarrow$ & SIM $\uparrow$ & SingMOS $\uparrow$ & Sheet $\uparrow$ \\
\midrule
GroundTruth & - & 0.089 & - & 4.450 & 4.237 & 0.208 & - & 4.569 & 3.815 \\
\midrule
StyleSinger & Score & 0.388 & 0.808 & 3.892 & 3.833 & - & - & - & - \\
TCSinger & Score & 0.227 & 0.822 & 3.939 & 4.009 & 0.460 & 0.729 & 3.652 & 3.685 \\
YingMusic-Singer & Melody & 0.117 & 0.913 & 4.068 & 3.599 & - & - & - & - \\
Vevosing & Melody & 0.233 & 0.908 & 4.265 & 3.814 & 0.256 & 0.888 & 4.184 & 3.582 \\
\midrule
SoulX-Singer & Melody & \textbf{0.069} & 0.902 & \textbf{4.394} & \textbf{4.053} & 0.155 & 0.870 & 4.223 & \textbf{3.703} \\
SoulX-Singer & Score & \textbf{0.069} & \textbf{0.922} & 4.370 & 4.010 & \textbf{0.129} & \textbf{0.914} & \textbf{4.255} & 3.690 \\
\bottomrule
\end{tabular}
}
\end{table}

\subsubsection{Cross-Lingual Synthesis Evaluation}

To further evaluate the model’s ability in cross-lingual synthesis (e.g., using a Mandarin prompt to synthesize English singing), which requires strict disentanglement of speaker identity from linguistic content, we conducted experiments reported in Table~\ref{tab:cross_lingual}. The baseline model Vevosing exhibits severe intelligibility degradation, with a WER as high as 0.717, indicating leakage of linguistic patterns from the prompt into the generated output. In contrast, SoulX-Singer achieves a WER of 0.110 while maintaining a high SIM of 0.898, demonstrating strong preservation of speaker identity across languages. These results highlight the effectiveness of the singing content encoder, which successfully separates language-independent timbre features from language-dependent linguistic content, enabling high-fidelity cross-lingual style transfer without compromising pronunciation or vocal identity.


\begin{table}[htbp]
\centering
\small
\caption{Cross-Lingual Synthesis Performance on SoulX-Singer-Eval ($\uparrow$ larger is better, $\downarrow$ smaller is better; bold indicates the best result).}
\label{tab:cross_lingual}
\vspace{2mm}
\begin{tabular*}{\linewidth}{@{\extracolsep{\fill}}llcccc}
\toprule
\multirow{2}{*}{Model} & \multirow{2}{*}{Control} & \multicolumn{4}{c}{\textbf{Cross-Lingual}} \\
\cmidrule(lr){3-6}
 & & WER $\downarrow$ & SIM $\uparrow$ & SingMOS $\uparrow$ & Sheet $\uparrow$ \\
\midrule
GroundTruth & - & 0.148 & - & 4.510 & 4.026 \\
\midrule
TCSinger & Score & 0.333 & 0.789 & 3.817 & 3.830 \\
Vevosing & Melody & 0.717 & 0.877 & 4.133 & 3.631 \\
\midrule
SoulX-Singer & Melody & 0.122 & 0.866 & \textbf{4.342} & \textbf{3.939} \\
SoulX-Singer & Score & \textbf{0.110} & \textbf{0.898} & 4.337 & 3.883 \\
\bottomrule
\end{tabular*}
\end{table}

\section{Conclusions}

In this work, we introduced SoulX-Singer, a high-fidelity, zero-shot singing voice synthesis model. Leveraging a note-level aligned dataset of over 42,000 hours, the model enables flexible control over both melody and score, while faithfully reproducing diverse timbres. Extensive evaluations across multiple benchmarks demonstrate that SoulX-Singer consistently outperforms state-of-the-art baselines in pitch accuracy, intelligibility, timbre similarity, and overall singing quality. By combining explicit acoustic and symbolic cues with large-scale data, SoulX-Singer provides a robust foundation for creative applications, including personalized singing synthesis, music production, and multilingual vocal content generation. This work paves the way for future research in zero-shot expressive SVS and singing voice editing.

\section{Ethics Statement}

The development and release of SoulX-Singer raise important ethical considerations. As a zero-shot singing voice synthesis system, the model is capable of generating realistic singing voices conditioned on a short reference prompt, which may introduce risks related to voice impersonation and misuse. Users of SoulX-Singer are therefore strongly encouraged to respect intellectual property, privacy, and personal consent when generating singing content. The system should not be used to impersonate individuals without authorization, nor to produce deceptive or misleading audio content.

\section{Acknowledgments}

We would like to express our sincere gratitude to Rizhao Youle Studio and the following individuals for contributing high-quality vocal recordings and providing valuable assistance in the construction of the SoulX-Singer-Eval benchmark: 
Zhengyu Chen, Hao Fan, Jialan Kuang, Haotian Luo, Linfei Ma, Hao Meng, Rui Shi, Wenbin Shi, Yucun Sui, Xuewen Sun, Jiayu Wang, Weiliang Wang, Shuhua Weng, Minghao Xu, Dan Yang, Baojin Yi, Youpeng Yuan, Meng Zhang, Peng Zhao, and Zhiyi Zheng.

\printbibliography

\newpage

\section*{Appendix}

\subsection*{SoulX-Singer-SVC}

To extend SoulX-Singer to scenarios where precise MIDI and time-aligned lyrics are unavailable, we further introduce SoulX-Singer-SVC, a singing voice conversion (SVC) model derived from SoulX-Singer. 
Specifically, we replace the original score and text encoders with a frozen Whisper-base encoder to extract robust semantic representations directly from the source singing audio. Meanwhile, the F0 encoder and the flow-matching decoder are initialized with pre-trained SoulX-Singer parameters to preserve melody controllability and high-fidelity synthesis quality. The model is then fine-tuned on a relatively small but high-quality singing voice dataset.

We evaluate SoulX-Singer-SVC on GMO-SVS and SoulX-Singer-Eval, comparing it with GroundTruth, melody-controlled SoulX-Singer, and state-of-the-art SVC systems, including YingMusic-SVC \cite{chen2025yingmusic} and Vevo-SVC \cite{zhang2025vevo2}. Note that the FFE metric is omitted for SoulX-Singer-Eval because the source audio was pitch-shifted. The results are presented in Table~\ref{tab:svc_overall}, and the model is released as part of the SoulX-Singer project.

\begin{table*}[htp]
\centering
\small
\caption{Singing voice conversion performance results on two datasets ($\uparrow$ larger is better, $\downarrow$ smaller is better). The best and the second best results are shown in \textbf{bold} and \underline{underlined}.}
\label{tab:svc_overall}
\vspace{2mm}
\resizebox{\linewidth}{!}{
\begin{tabular}{lcccccccccc}
\toprule

\multicolumn{11}{c}{\textbf{GMO-SVS Dataset}} \\
\midrule

\multirow{2}{*}{Model} & \multicolumn{5}{c}{Chinese} & \multicolumn{5}{c}{English} \\
\cmidrule(lr){2-6} \cmidrule(lr){7-11}
& WER$\downarrow$ & SIM$\uparrow$ & FFE$\downarrow$ & SingMOS$\uparrow$ & Sheet$\uparrow$
& WER$\downarrow$ & SIM$\uparrow$ & FFE$\downarrow$ & SingMOS$\uparrow$ & Sheet$\uparrow$ \\

\midrule

GroundTruth & 0.074 & - & - & 4.624 & 4.334 & 0.197 & - & - & 4.441 & 3.825 \\
SoulX-Singer (Melody) & 0.065 & 0.897 & 0.044 & 4.458 & 4.110 & 0.151 & 0.918 & 0.036 & 4.323 & 3.751 \\

\midrule
YingMusic-SVC & \textbf{0.097} & \textbf{0.908} & \underline{0.057} & 4.399 & 3.974 & \textbf{0.243} & \textbf{0.923} & \underline{0.046} & \underline{4.337} & \underline{3.708} \\
Vevo-SVC & 0.143 & 0.892 & 0.104 & \textbf{4.471} & \textbf{4.088} & 0.334 & \underline{0.920} & 0.079 & \textbf{4.389} & \textbf{3.748} \\
SoulX-Singer-SVC & \underline{0.107} & \underline{0.904} & \textbf{0.036} & \underline{4.419} & \underline{4.029} & \underline{0.267} & 0.919 & \textbf{0.030} & 4.314 & 3.666 \\

\midrule
\multicolumn{11}{c}{\textbf{SoulX-Singer-Eval Dataset}} \\
\midrule

\multirow{2}{*}{Model} & \multicolumn{5}{c}{Chinese} & \multicolumn{5}{c}{English} \\
\cmidrule(lr){2-6} \cmidrule(lr){7-11}
& WER$\downarrow$ & SIM$\uparrow$ & FFE$\downarrow$ & SingMOS$\uparrow$ & Sheet$\uparrow$
& WER$\downarrow$ & SIM$\uparrow$ & FFE$\downarrow$ & SingMOS$\uparrow$ & Sheet$\uparrow$ \\

\midrule

GroundTruth & 0.089 & - & - & 4.450 & 4.237 & 0.208 & - & - & 4.569 & 3.815 \\
SoulX-Singer (Melody) & 0.072 & 0.869 & - & 4.402 & 4.061 & 0.161 & 0.883 & - & 4.249 & 3.755 \\

\midrule
YingMusic-SVC & \textbf{0.096} & \underline{0.876} & - & 4.227 & 3.848 & \textbf{0.236} & 0.881 & - & 4.274 & 3.727 \\
Vevo-SVC & 0.180 & 0.871 & - & \textbf{4.387} & \underline{3.932} & 0.346 & \underline{0.886} & - & \textbf{4.359} & \underline{3.772} \\
SoulX-Singer-SVC & \underline{0.102} & \textbf{0.880} & - & \underline{4.303} & \textbf{3.938} & \underline{0.260} & \textbf{0.894} & - & \underline{4.286} & \textbf{3.774} \\

\bottomrule
\end{tabular}
}
\end{table*}

\end{document}